# Field-free Magnetization Switching by Utilizing the Spin Hall Effect and Interlayer Exchange Coupling of Iridium

Yang Liu[1,3], Bing Zhou[1,3], and Jian-Gang (Jimmy) Zhu*[2,3]

*1 Department of Materials Science and Engineering, Carnegie Mellon University, 5000 Forbes Avenue, Pittsburgh, Pennsylvania, 15213, USA*

*2 Department of Electrical and Computer Engineering, Carnegie Mellon University, 5000 Forbes Avenue, Pittsburgh, Pennsylvania, 15213, USA*

*3 Data Storage Systems Center, Carnegie Mellon University, 5000 Forbes Avenue, Pittsburgh, Pennsylvania, 15213, USA*

**Abstract**

Magnetization switching by spin-orbit torque (SOT) via spin Hall effect represents as a competitive alternative to that by spin-transfer torque (STT) used for magnetoresistive random access memory (MRAM), as it doesn't require high-density current to go through the tunnel junction. For perpendicular MRAM, however, SOT driven switching of the free layer requires an external in-plane field, which poses limitation for viability in practical applications. Here we demonstrate field-free magnetization switching of a perpendicular magnet by utilizing an Iridium (Ir) layer. The Ir layer not only provides SOTs via spin Hall effect, but also induce interlayer exchange coupling with an in-plane magnetic layer that eliminates the need for the external field. Such dual functions of the Ir layer allows future build-up of magnetoresistive stacks for memory and logic applications. Experimental observations show that the SOT driven field-free magnetization reversal is characterized as domain nucleation and expansion. Micromagnetic modeling is carried out to provide in-depth understanding of the perpendicular magnetization reversal process in the presence of an in-plane exchange coupling field.

* Corresponding author, jzhu@cmu.edu

Research efforts in discovering and gaining better understanding of various spin-based physical phenomena over the past decades have propelled the innovation and developments of new generations of memory and logic devices[1-4]. With utilization of non-volatility inherent in magnetism and low-power consumption characteristics, these novel device concepts present new opportunities for future electronics and computers. One of the most renowned examples is the spin-transfer torque magnetoresistive random access memory (STT-MRAM)[5-8]. In STT-MRAM, spin polarized current can be generated and used to switch the magnetization direction via spin momentum transfer. In recent years, a three-terminal scheme for MRAM applications has been proposed that utilizes spin-orbit torques (SOTs) to switch the free layer[9-11]. Such scheme shows the benefits in lower current density as well as separate read- and write-path, and thus has attracted people's strong interests. In these devices, spin Hall effect (SHE)[12,13] is adopted as the main source for SOT injection[14-16]. It has been demonstrated that SHE can trigger magnetization switching of in-plane magnetic tunnel junctions (MTJs)[15,17,18]. For perpendicular MTJs that are required for high-density memory, an in-plane magnetic field has to be applied for SHE induced switching[16,18]. However, the need for an external applied field significantly hinders the technological viability of commercial applications.

By far, people have come up with several ways to solve this problem. One approach is by creating geometric asymmetry so that the SOTs acting on the perpendicular magnet could contain perpendicular components[19-21]. But this approach suffers the scalability issue. Another way is by introducing an intrinsic in-plane field by exchange bias or interlayer exchange coupling. It has been reported that an antiferromagnetic layer, such as PtMn[22] and IrMn[23,24], is able to generate spin current as well as add a horizontal field to the neighboring perpendicular magnetized layer by exchange bias so as to assist the switching. The detailed mechanisms, however, is not fully understood including SHE of antiferromagnetic materials as well as the large slopes in the switching loops. Moreover, YC Lau et al. obtained field-free magnetization reversal by using the interlayer exchange coupling via Ru[25]. But one limitation of this approach is that one cannot further add magnetoresistive layers to the film stack because the exchange coupling layer and spin Hall layer are separated.

Here, we demonstrate the realization of field-free magnetization switching by utilizing an Ir interlayer layer. Ir has been shown capable of generating spin Hall effect[26,27], and equally important for this study, it provides interlayer exchange coupling when sandwiched by two ferromagnetic layers, e.g. Co[28,29] at adequate thicknesses. Combing these two properties, we are able to achieve deterministic magnetization switching without an external in-plane field. We also investigate the switching process in our devices and characterize it as domain nucleation followed with thermally assisted SHE-induced domain wall motion (DWM).

**Results and discussion**

The film stack for field-free magnetization switching is substrate/ Co(2)/ Ru(0.85)/ Co(2)/ Ir(1.35)/ Co(1.2)/ Ta(2), with unit in nanometers. Here, the top thin Co layer is naturally perpendicularly magnetized due to the interfacial perpendicular anisotropy arising from the interface with the Ir layer[28,29]. The thickness of the Ir layer corresponds to the second antiferromagnetic coupling peak in the Ruderman–Kittel–Kasuya–Yosida (RKKY) thickness dependence curve (see supplementary information Fig. S1). The Co/Ru/Co below the Ir layer is a flux-matched in-plane synthetic antiferromagnetic (SAF) structure. The in-plane hysteresis loop of the SAF (Fig. 1(b)) shows strong interlayer exchange coupling and the loop shape indicates the single domain characteristics for the in-plane Co layers. The fabricated film is patterned into Hall-cross devices with 4-µm wide current channel and 1-µm wide voltage channel, as shown in Fig. 1(c). Fig. 1(d) shows the measured Hall voltage arising from anomalous Hall effect (AHE) with varying external perpendicular fields. The squared shape of the AHE loop indicates well defined perpendicular magnetic anisotropy of the top Co layer.

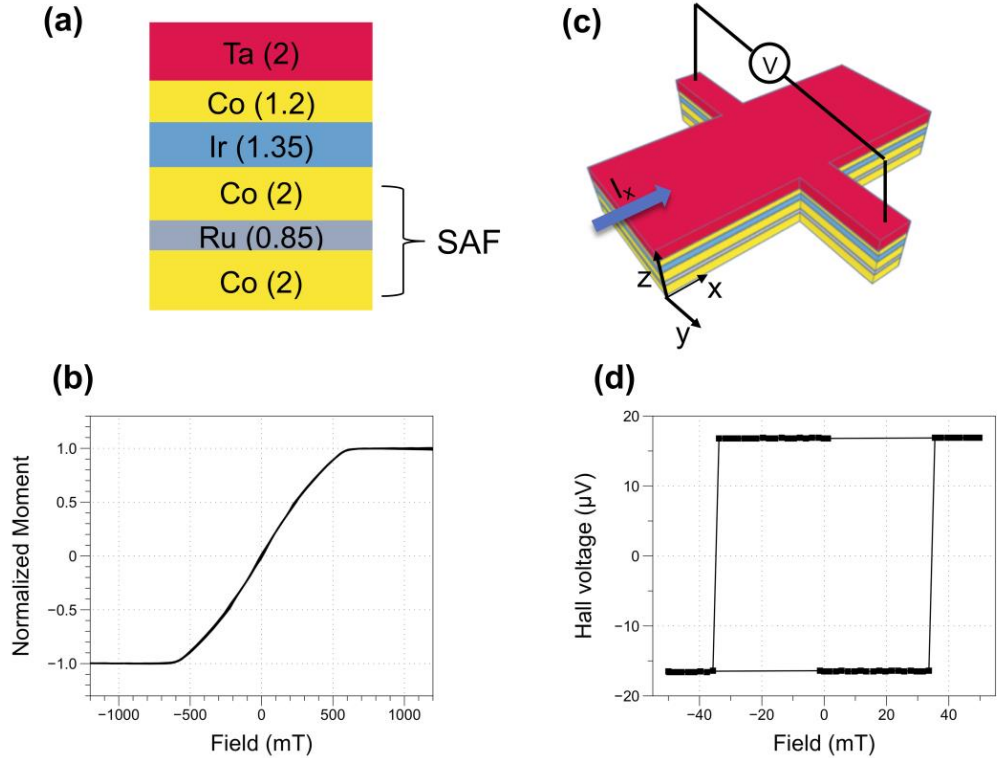

FIG. 1. (a) Film stack for field-free magnetization switching, with unit in nanometers. (b) In-plane hysteresis loop of the bottom SAF. (c) Schematic illustration of the Hall-bar device for spin-orbit torque measurement. (d) Anomalous Hall effect loop showing the perpendicular magnetic anisotropy of the top Co layer

To study the SOT-driven magnetization switching, current pulses are injected along the current channel with each pulse length of 100 µs. The Hall voltage was measured after each write pulse with a read current of 100 µA, less than 1/10 of the writing current. Fig. 2(a) shows the anomalous Hall voltage as a function of current density with a series of external fields, $H_x$, applied along the current direction. For $H_x > 20$ mT, well defined switching loops are obtained with positive and negative saturation values corresponding to the two perpendicular magnetization states. At $H_x = 20$ mT, the switching hysteresis loop collapses, indicating no magnetization switching occurs. At zero external field, the switching completely recovers, however, with flipped loop shape. The same switching loops are maintained for $H_x < 0$. Fig. 2(b) shows the similar measurement sequence with the reversed magnetization of the bottom SAF structure (from ⇄ to ⇆). In contrast to that shown in Fig. 2(a), the collapse of the switching hysteresis loop now occurs at $H_x = -20$ mT instead.

It is our interpretation that when charge current flows into the Ir layer, pure spin current is generated due to spin Hall effect and then injected to the top perpendicular Co layer. This pure spin current drives the observed magnetization reversals provided there exists an in-plane magnetic field. At zero applied field, this in-plane field arises from the interlayer coupling between the perpendicular Co and in-plane Co layers above and below the Ir interlayer via the RKKY interaction. When the externally applied field cancels the coupling field in magnitude and direction exactly, magnetization reversal no longer occurs. Our evidence suggests this in-plane coupling field has a magnitude of 20 mT. It should be noted that the 20 mT coupling field shown in the devices is significantly smaller than that measured at film level, which is around 95 mT.

The above argument is further confirmed by the measurements on the devices that have the film stack without the bottom SAF structure (see Supplementary information Fig. S2). AHE hysteresis loops resulting from perpendicular magnetization switching can only be obtained with external in-plane field whereas no current-induced switching is observed at zero field.

It's noteworthy that the use of the SAF structure eliminates the possibility of stray field effect since the two antiparallel magnetized Co layers in the SAF is essentially flux-matched.

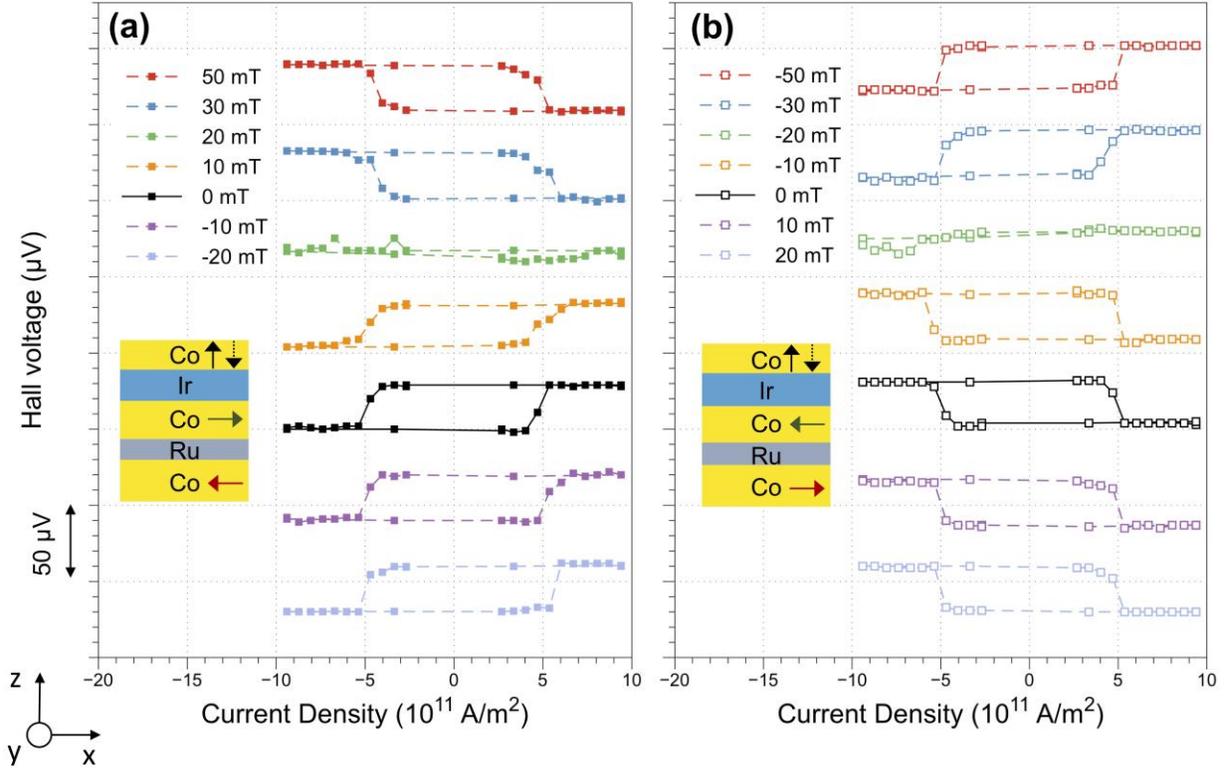

FIG. 2. Anomalous Hall voltage as a function of injected current density in the Ir layer with various external magnetic fields $H_x$ along the current direction. Bottom SAF is set as (a) ⇄, and (b) ⇆

To understand the nature of current-induced magnetization switching in the presence of the in-plane coupling field in our devices, Kerr microscopy is utilized to visualize magnetization reversal process, as shown in Fig. 3, with current pulses of 100 ns in duration. It's observed that small opposite domains nucleate at the beginning of the reversal. The nucleation sites mostly locate around the device edge which is likely due to the degradation of perpendicular anisotropy possibly caused by the etching process during device fabrication. Continued application of subsequent current pulses results in the expansion of the reversed domains in all directions. With sufficient number of current pulses, the region along the current pulses can be mostly reversed. We should note that there are always some small residual domains located sparsely along the current path that are difficult to eliminate, even with perpendicular magnetic field.

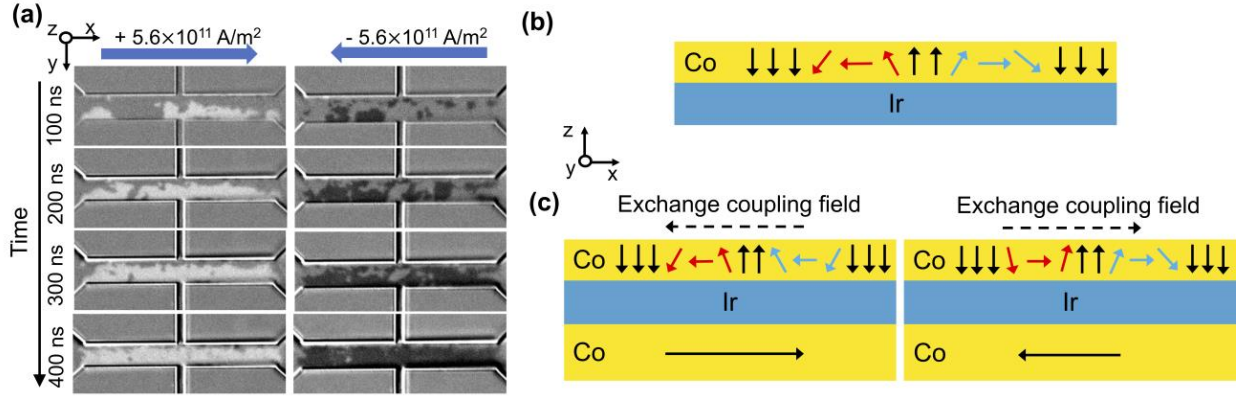

FIG. 3 (a) Kerr microscope images showing the domain wall propagation during the current-induced magnetization switching process, without external magnetic field. Current is applied in +x (left) and -x (right) directions. Bottom SAF is set as ⇄. Between images in each column: 100-ns pulse, 1 pulse. (b) Schematic illustration of right-handed domain wall structures in Ir/Co system. Red arrows: down-up domain wall. Blue arrows: up-down domain wall. (c) Domain wall structures in presence of the exchange coupling field.

It's widely observed that SHE can induce domain wall motion (DWM) in heavy metal/ ferromagnet systems[30–34]. One of the keys to the SHE-driven DWM is the Dzyaloshinskii-Moriya interaction (DMI) at heavy metal/ ferromagnet that not only makes Neel-type domain walls more energetically favorable than Bloch-type ones but also introduces chirality to the domain wall structure[35–37]. Normally, up-down and down-up domain walls of an opposite domain possess the same chirality (either left- or right-handed), as shown in Fig. 3(b). These domain walls move in the same direction when receiving the SOTs from the neighboring heavy metal layer and thus domains can only shift forwards or backwards. With applying external in-plane field that's larger than the effective DMI field, up-down and down-up domain walls can have different chiral structures. As a result, a domain can either expand or shrink depending on the direction of injected current[38], which further leads to deterministic magnetization switching. In our case, we measured the effective DMI field at Ir/Co interface to be about 9 mT (see supplementary information Fig. S4). Hence, the exchange coupling field (20 mT) in our devices is able to overcome the effective

DMI field and give rise to the domain walls with opposite chirality (Fig. 3(c)). Based on our observations, it's the two types of domain walls moving in opposite directions that enables field-free magnetization switching. Therefore, the magnetization switching process in our devices can be interpreted as domain nucleation followed with thermally assisted SHE-driven DWM till the expansion of reversed domains produces full reversal.

Recently, CB Seung-heon, *et al.*[39] reported that spin current generated by the interface of non-magnetic metal/ in-plane magnetized ferromagnet can contain some perpendicular polarization. Such perpendicular component of the spin current further gives rise to the SOTs that can achieve field-free magnetization switching. This mechanism can't be applied to explain our observations since the spin diffusion length of Ir is only 0.5 nm[40]. Therefore, even if such spin current is generated at Ir/ in-plane Co interface it can't travel through the 1.35 nm-thick Ir layer to reach the top perpendicular Co layer.

To provide further understanding of the experimental observations, the effect of pure spin current in the perpendicular Co layer is modeled by Slonczewski spin transfer torque included Landau-Lifshitz-Gilbert damped gyromagnetic equation of motion. The full stack is modeled with the bottom SAF structure set as ⇄. The modeling is carried out by initially creating a reversed domain, or domains, in a perpendicularly saturated top Co layer. Fig. 4 shows sets of time evolution of domain configurations during the expansion of the initial reverse domains under the pure spin current. It is found that the reversed domains expand in all directions. Domain walls that are parallel to the interlayer exchange coupling field (as in the top left) move as fast as those orthogonal to the field (top right). It is interesting to note that as the reverse domains expand, the wall fronts often become curved during their motion (except when the wall front is perpendicular to the exchange field, as the case of top right in the Fig.4 shows.). This is because the wall moving speed is correlated with the orientation of magnetization component at the center of the wall. Fig. 5 shows the wall speed as a function of the along-wall projection of the magnetization at the center of the wall with various interlayer exchange coupling strength. The simulation results suggest that the wall moves faster when the magnetization at the center of the wall becomes more parallel to the tangent of the wall. Furthermore, it can also be seen that not only the speed of DWM but also the slope of the speed angular dependence increase with higher interlayer exchange coupling field.

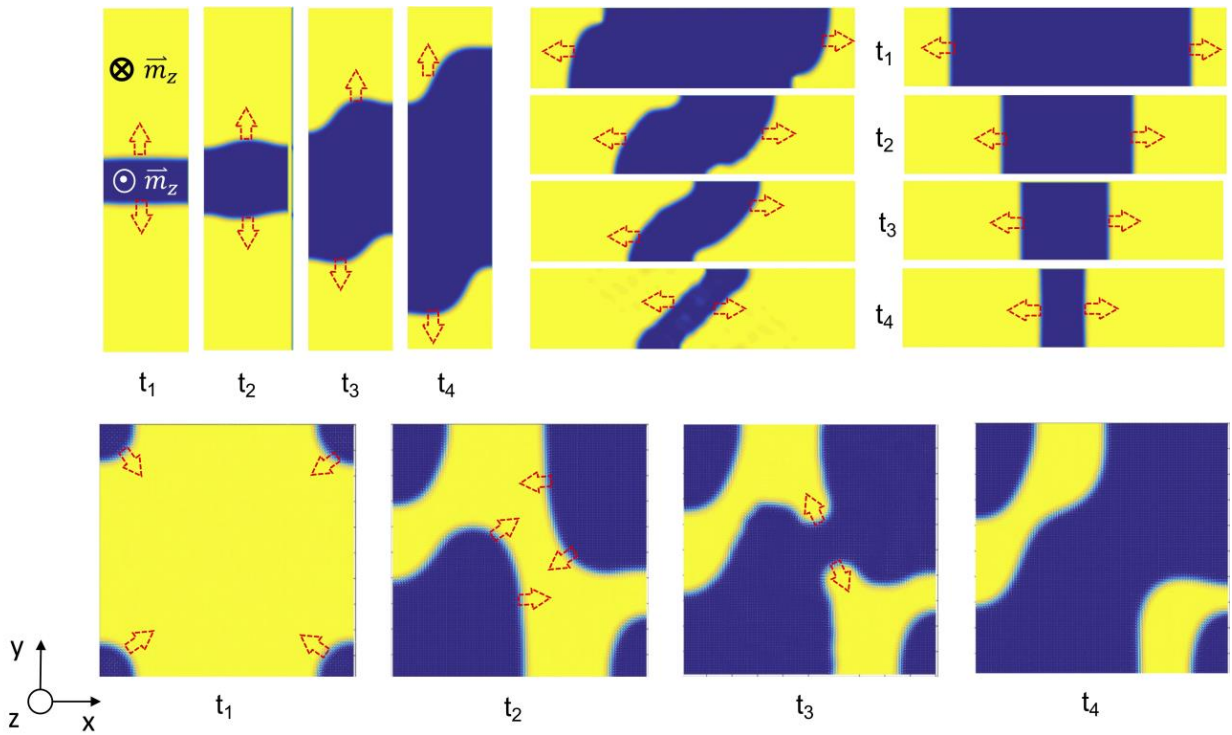

FIG. 4. Simulated domain expansion in the perpendicular Co layer in the presence of the in-plane coupling field. The electron charge current flows along the x-axis. $t_1 < t_2 < t_3 < t_4$ represents the time evolution during the application of current.

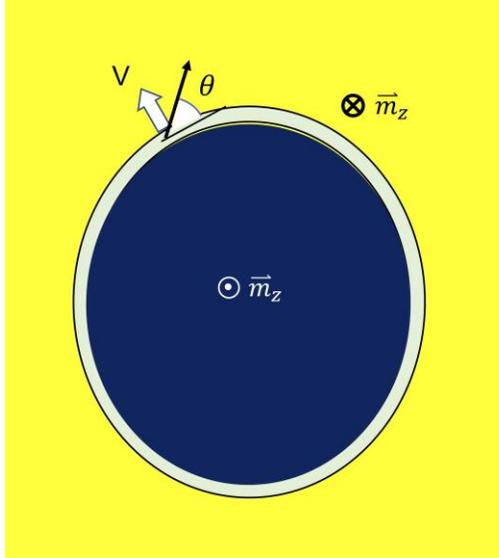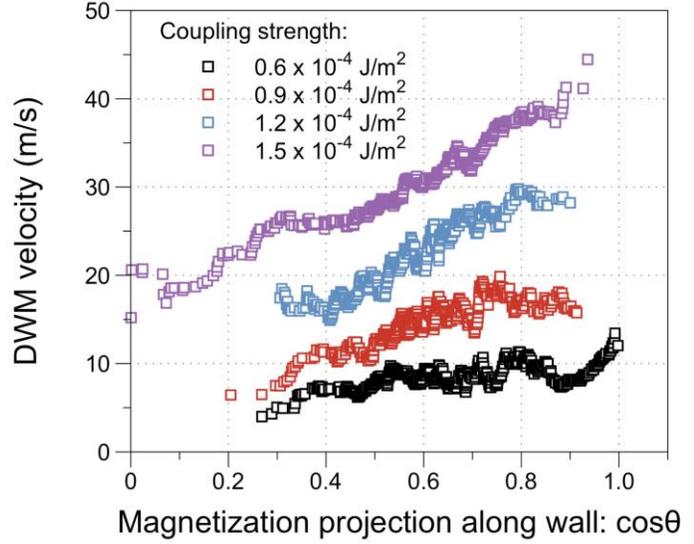

FIG. 5. Calculated domain wall moving speed under the SHE generated spin current as a function of the direction cosine of the wall center magnetization along the wall tangent with four different in-plane coupling field strengths.

**Conclusion**

To summarize, we achieved robust field-free magnetization switching by utilizing the SHE and interlayer exchange coupling of Ir. The switching process is dominated by SHE-induced domain wall propagation to achieve the full expansion of reserved domains. Combined modeling study shows that in the presence of the in-plane coupling field, the nucleated domains can expand in all directions and the higher the coupling field, the higher the expansion speed of the reversed domains. The domain wall speed also increases when the magnetization at the center of the wall becomes more parallel to the wall during the motion. It's noteworthy that the film stack we used is easy to be built up with magnetoresistive layers on the top. Meanwhile, the device size should be able to scale down to the dimension of tens of nanometers, which shows its potential for practical applications in memory and logic.

**Method**

All films are deposited at room temperature by magnetron sputtering with base pressure < $2 \times 10^{-8}$ Torr. Film-level hysteresis loops are measured by Alternating Gradient Field Magnetometer (AGFM). The films are patterned into Hall-cross devices with 4-$\mu$m wide current channel and 1-$\mu$m wide voltage channel utilizing e-beam lithography, optical lithography and ion beam etching. To study the SOT-driven magnetization switching, current pulses are injected along the current channel with each pulse length of 100 $\mu$s. The amplitude of the current at milliamp (mA) level is swept from negative to positive values and back to negative again. The Hall voltage was measured after each write pulse with a read current of 100 $\mu$A. Prior to writing, the moments of bottom SAF is initially set as ⇄ (or ⇆) by applying 1 T magnetic field in +x (or -x) direction. The current density is calculated with taking into the account both the resistivity and cross-section area of each layer. For studying the domain wall propagation during magnetization switching process, 100 $n$s current pulses are injected and the dynamics of magnetization reversal was tracked by Kerr microscope.

For DW modeling in our devices, the interlayer exchange coupling strength between the perpendicular Co layer and the in-plane Co layer sandwiching the Ir layer is varied throughout the study. The top perpendicular Co layer is assumed to have 2 nm thickness with interfacial perpendicular uniaxial anisotropy of Ks = 4 x $10^{-3}$ J/m$^2$ and saturation magnetization Ms = 1.2 T. The two Co layers in the Co/Ru/Co SAF are assumed to be flux-matched without perpendicular anisotropy and to have antiparallel exchange coupling of $\sigma_{ex,SAF}$ = $-1.0\times10^{-3}$ J/m$^2$ . The film stack is meshed laterally with each mesh cell of 2 x 2 nm$^2$ size. The exchange stiffness constant A = 1.6 x $10^{-11}$ J/m is assumed for all three Co layers. A spatially uniform pure spin current with spin polarization along the positive y-direction is assumed corresponding to a charge current density of $J_c = 10^{12}$ A/m² and a spin Hall angle of 10%. The top Co layer of the SAF is always magnetized along the positive x-direction with essentially uniform magnetization. Zero DMI is assumed for simplification.

**Acknowledgments**

This work was funded by a Laboratory Directed Research and Development grant from Sandia National Laboratories. Sandia National Laboratories is a multi-mission laboratory managed and operated by National Technology and Engineering Solutions of Sandia, LLC., a wholly owned subsidiary of Honeywell International, Inc., for the U.S. Department of Energy's National Nuclear Security Administration under contract DE-NA0003525. The authors thank Calvin Chan and Alex Roesler for their support.

Supplementary Information

## 1. Interlayer exchange coupling of Iridium (Ir)

Film stacks for investigating the interlayer exchange coupling of Ir are Substrate/ Co (2 nm)/ Ir ($t_{Ir}$)/ Co (2 nm)/ Ta (2 nm) where $t_{Ir}$ ranges from 0.6 nm to 1.65 nm. In-plane hysteresis loops were measured by alternating gradient field magnetometer (AGFM) and used to determine the strength of interlayer exchange coupling. Figure S1 shows the exchange coupling field as a function of Ir layer thickness. Note that only antiferromagnetic coupling is observed in our experiments. As can be seen, the first antiferromagnetic coupling peak locates at 0.6 nm and the second at 1.35 nm. For the magnetization switching experiments, we choose the Ir layer thickness to be 1.35 nm for that the exchange coupling via 0.6 nm Ir is so strong that the perpendicular Co layer can be pulled into in-plane direction by its coupled in-plane Co layer.

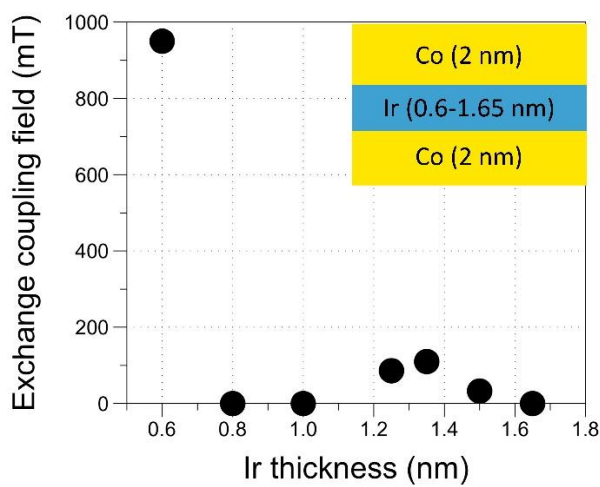

FIG. S1. Exchange coupling field as a function of Ir layer thickness

## 2. Spin Hall effect (SHE) induced magnetization switching in the device without the in-plane magnetized layers.

We also fabricated the devices that are of the film stack without the bottom in-plane magnetized layers, that is Substrate/ Ta (0.5 nm)/ Ir (3 nm)/ Co (1.2nm)/ Ta (2nm). SHE induced magnetization switching measurements were conducted and the results are shown in Figure S2. In this case, no observable switching is obtained in absence of an external field. On the other hand, magnetization reversal is observed with external field $H_x = \pm 10$ mT. The switching loops are flipped when reversing the in-plane field direction. This is a typical behavior of SHE induced switching for that the preferred magnetization state depends on the in-plane field direction.

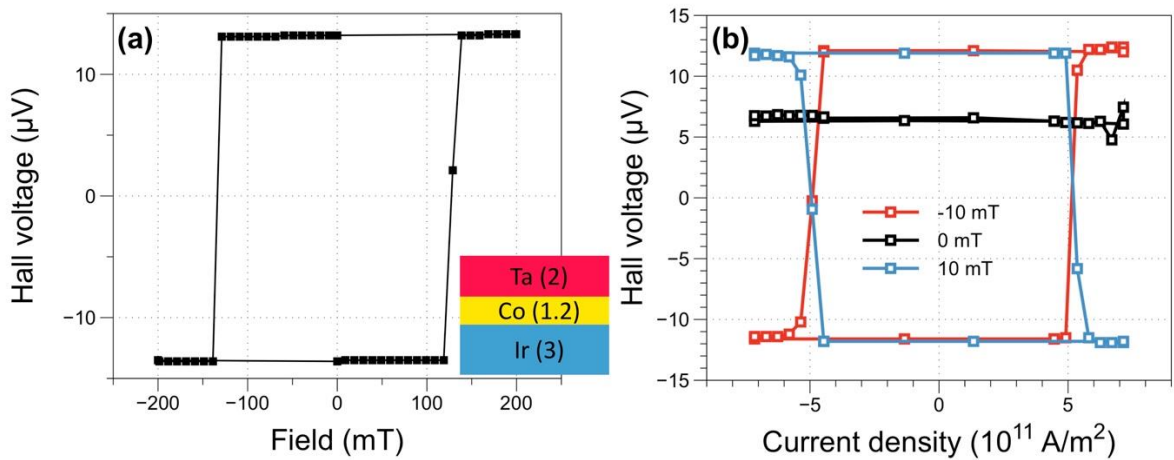

FIG. S2. (a) Anomalous Hall effect loop of the device without the bottom in-plane Co layers. (b) Current induced magnetization switching under different external fields.

## 3. Spin Hall effect (SHE) induced magnetization switching in the device with a thick Ir layer.

Now the stack contains a 3 nm-thick Ir layer, as shown in Figure S3. With this thickness, the strength of interlayer exchange coupling should decay to nearly zero. Results shows that in this case magnetization reversal can only occur with applying an external field $H_x$. This partially proves the field-free magnetization switching presented in the main body of manuscript is facilitated by the in-plane field resulting from the interlayer exchange coupling via the 1.35 nm Ir.

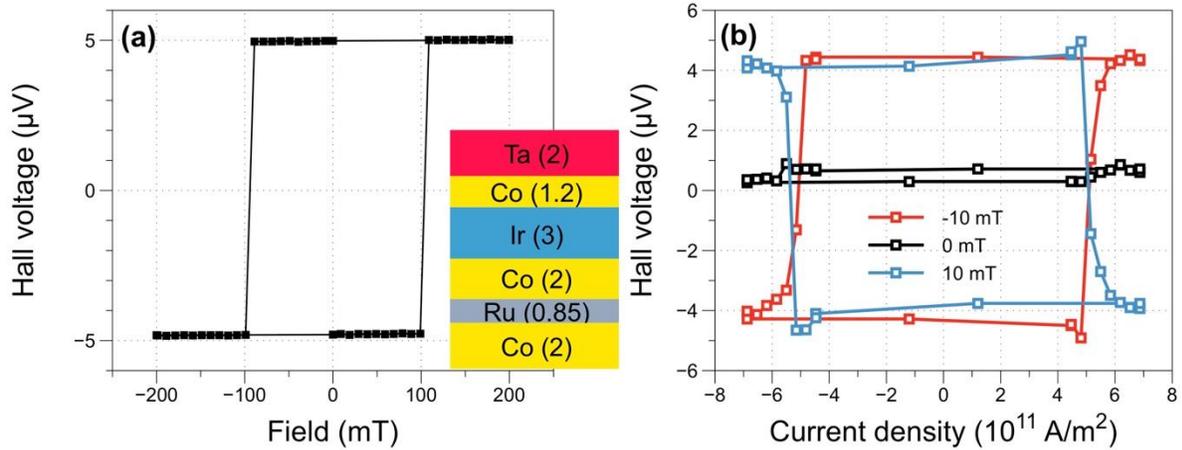

FIG. S3. (a) Anomalous Hall effect loop of the device with a 3 nm Ir layer. (b) Current induced magnetization switching under different external fields.

## 4. Domain wall chirality and effective DMI field at Ir/Co interface

To characterize the domain wall chirality as well as to measure the effective DMI field at Ir/Co interface, the growth of magnetic bubble domains was measured with Kerr microscope. During this measurement, a magnetic bubble domain was first initiated. Then a static in-plane field $H_x$ was applied whereas perpendicular field ($H_z = 10$ mT) pulses were generated to grow the bubble domain. The growing dynamics was recorded and used for analyzing the growth velocity of up-down and down-up domain walls. A summary of domain wall velocity as a function of in-plane field is plotted as Figure S4. Typically, higher domain wall velocity is favored when $H_x$ is parallel to the direction of effective DMI field within the domain wall. Thus, our observations show that the domain wall in our films is right-handed Neel-type wall. The two minima in domain wall velocity plot indicates the effective DMI field is balanced with $H_x$ when they are antiparallel.

It implies that the effective DMI field at Ir/Co interface is around 9 mT, much smaller than that at Pt/Co and W/FeCoB interface.

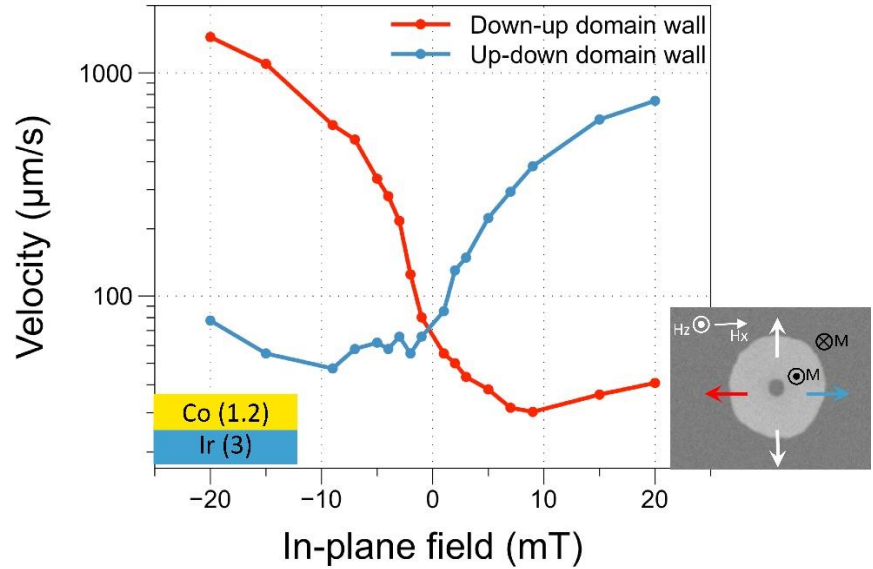

FIG. S4. Velocity of down-up and up-down domain wall as a function of in-plane magnetic field. This plot shows the chirality of domain wall at Ir/Co interface is right-handed. The effective DMI field is about 9 mT.